\documentclass[12pt]{article}

\usepackage[T1]{fontenc}
\usepackage[utf8]{inputenc}
\usepackage{lmodern}
\usepackage{microtype}
\usepackage{float}
\usepackage{changepage}
\usepackage{tabularx}

\usepackage{geometry}
\usepackage{amsmath, amssymb, amsthm}

\usepackage{csquotes}

\usepackage{graphicx}
\usepackage{booktabs}
\usepackage[font=small,labelfont=bf,labelsep=period]{caption}

\usepackage{enumitem}

\usepackage{titlesec}

\titleformat{\section}
  {\large\bfseries\centering}
  {\thesection}
  {0.75em}
  {}

\titleformat{\subsection}
  {\normalsize\bfseries\centering}
  {\thesubsection}
  {0.75em}
  {}

\titlespacing*{\section}{0pt}{2.6em}{1.2em}

\usepackage[colorlinks=true,
            linkcolor=black,
            citecolor=black,
            urlcolor=black]{hyperref}

\title{
\LARGE Beyond Hurwicz\\[0.75ex]
\Large Incentive Compatibility under Informational Decentralization
}
\author{\normalsize David Lancashire}
\date{\normalsize January 2026}

\newenvironment{narrowfigure}
  {\begin{center}\begin{minipage}{0.8\textwidth}}
  {\end{minipage}\end{center}}

\begin{document}

\maketitle
\vspace{1.8em}

\begin{center}
\begin{minipage}{0.75\textwidth}
\begin{abstract}
\small
Achieving incentive compatibility under informational decentralization is impossible within the class of direct and revelation-equivalent mechanisms typically studied in economics and computer science. We show that these impossibility results are conditional by identifying a narrow class of non-revelation-equivalent mechanisms that sustain enforcement by inferring preferences indirectly through parallel, uncorrelatable games.
\end{abstract}
\end{minipage}
\end{center}

\vspace{1.5em}

\section{Introduction}

If we read the \emph{Churning of the Milk Ocean} as a Hindu creation myth, what
surprises is not so much that a price must be paid for setting the world in
motion, but that the cost cannot be borne by the gods acting in the field of
creation. Before time allows its deeper treasures to emerge, external authority
is needed --- Shiva --- to stabilize the system.

Hurwicz echoes this structural constraint in his foundational papers on
implementation theory, arguing that whenever participants can mislead others
about their preferences, incentive compatibility requires an external authority
to punish deviations. Hurwicz saw this as a fundamental impossibility
result, and concluded it made incentive compatibility impossible in all
informationally decentralized systems lacking authorities to enforce transfers.

\begin{quote}\small
These results show that the difficulty is due not to our lack of inventiveness, but to a fundamental conflict among such mechanism attributes as the optimality of equilibria, incentive\-compatibility of the rules, and the requirements of informational decentralization.\footnote{Hurwicz, L. (1972). On the design of mechanisms for resource allocation. \emph{American Economic Review Papers \& Proceedings}, 62(2), 1-30.}
\end{quote}

Designers have accepted these arguments rather than confronting them directly.
Where auctions have required protection against agents ``changing their minds'',
designers have generally relied on courts, auctioneers, or institutions to
verify and punish attempts at revision. Such external authorities do more than
levy penalties: they determine the \emph{num\'eraire} in which agents are
penalized, forcing stabilizing punishments outside the scope of the mechanism as
necessary and creating a cost-bearing constraint that agents cannot neutralize
from within it.

There is, however, a narrow class of non-revelation-equivalent indirect mechanisms in which enforcement
costs can be generated endogenously. In these designs, agents who attempt to shift
equilibrium strategically must bear exposure to uncertainty over time, and this exposure
cannot be neutralized from within the mechanism. These mechanisms lie just
beyond the boundary identified by Hurwicz.

This paper explains how such solutions are possible. It starts with theory: why
circular mechanisms are unstable, and how designers stabilize them by
introducing enforcement costs that do not respond to strategic play. It then
shows it is possible to make these constraints responsive to play without making
deviation rational --- through design techniques which force agents to pay
front-loaded costs under uncertainty when proposing changes that will shift
equilibrium security levels in the mechanism. This permits incentive
compatibility to emerge in ways that are unimplementable in direct or
revelation-equivalent mechanisms.

\section{A Special Class of Circular Mechanisms}

In the canonical definition provided by Hurwicz (1972), a mechanism is simply a
function that maps actions or messages to outcomes. Whether a mechanism is
incentive compatible depends on whether its outcomes implement a social choice
rule as defined by Arrow (1951), and later Gibbard and Satterthwaite
(1973, 1975) and Maskin (1977).

It would take almost a decade before Myerson showed that under certain conditions
indirect mechanisms can be reduced to direct ones without loss of generality
(1979)\cite{Myerson1979}. The sheer power of his Revelation Principle would later induce many to
see indirect mechanisms as essentially interchangeable with direct ones. But in
the early 1970s, this suspicion had yet to take root, and Hurwicz framed his
definitions to cover games in both classes. Nor did he presuppose any particular
environment or specify how messages be communicated.

Rather, Hurwicz defined a mechanism as incentive compatible when, given the rules of the
game, each agent finds it optimal to act in a way that implements the intended
social choice rule. The circular mechanisms studied here fall squarely within
this definition. They map actions into outcomes. What distinguishes them from other mechanisms 
is a structural property: they operate in rounds with the results of early stages reappearing as constraints on later play.

For this reason, circular mechanisms should not be confused with repeated games that
extend over time. While circular mechanisms share similarities with such games,
what makes them uniquely vulnerable is that their own enforcement penalities -- the
costs that punish deviations -- are strategically responsive to play
within the game. Agents who learn how the game handles enforcement can
understand how to weaken it to enable deviation at lower cost.

Such mechanisms appear across many domains: as continuous double auctions and
adaptive market-clearing processes in economics; as reputation systems and
agenda-controlled games in political science; and as gossip protocols, adaptive
routing, and consensus mechanisms in computer science. Appendix~A provides a
representative taxonomy.

Existing research covers these games as well. Because their incentives are
path-dependent, we know players evaluate strategies over histories rather
than in isolation (Abreu, Pearce \& Stacchetti; Fudenberg \& Levine; Fudenberg \& Maskin)\cite{AbreuPearceStacchetti,FudenbergLevine1994,FudenbergMaskin1986}. Players may also reassess earlier choices in a phenomenon known as
\emph{time inconsistency} (see Strotz; Pollack; Hicks; Kydland \& Prescott), which
is destabilizing not only because agents may want to revise plans, but because
the possibility of revision may be anticipated (Hart \& Tirole;
Dewatripont; Barro \& Gordon). Incentive compatibility can fail even if all
parties regard the unrevised outcome as mutually beneficial (Farrell \& Maskin; Maskin \& Moore)\cite{FarrellMaskin1989,MaskinMoore1988}.

The problem with \emph{ex post revision} is that it creates the same need for a credible punishment
that Hurwicz identified in static settings. Hurwicz assumed such punishment was impossible in
decentralized environments lacking institutions capable of observing and
punishing inconsistency. His logic extends naturally to dynamic games with
incentives for revision, making it surprising that circular mechanisms can
eliminate this problem.

Yet the solution is oddly anticipated in a subset of theoretical papers. The most
striking is James Jordan’s discovery that when incentive
compatibility is unattainable in one-shot mechanisms, it can re-emerge in
path-dependent indirect mechanisms.\cite{Jordan1982a,Jordan1982b}
 Jordan does not argue that the solution
requires circularity, but demonstrates that any solution must distribute costs
over time. He also anticipates later work showing that direct mechanisms
can fail to implement equilibria available to indirect ones (Aumann; Renou \& Tomala; Strack \& Mora; Attar)\cite{Aumann1974,RenouTomala2012,StrackMora,Attar2021}, and that incentive compatibility can be restored
by leveraging expectations of future payoffs in their
enforcement structure (Maskin \& Moore; Abreu, Pearce \& Stacchetti).

The remainder of this paper shows how a narrow subclass of mechanisms can fulfill
these criteria by creating a dual-enforcement regime that allows the
mechanism to optimize its own security function by observing it indirectly in
the revealed preference players show to coordinate beyond the scope of its
enforcement guarantees.

\section{Endogenous Unactionability}

Sifting through paper after paper, we can see that where incentive compatibility
survives informational decentralization, it is because the mechanism embeds a
constraint that agents must bear but cannot neutralize. For the rest of this
paper, we refer to this property as \emph{endogenous unactionability}.

In mechanisms where external authorities punish deviation, endogenous
unactionability emerges trivially, as enforcers have no utility function in the
mechanism. Authorities must still be immune to strategic manipulation, but this
is handled by requiring them to verify any public states used to determine
allocations or transfers. This explains one of the reasons circular
mechanisms can appear counterintuitive: opacity -- not transparency -- is
what enables enforcement.

In computer science, many decentralized protocols exist which implement outcomes, 
and some accomplish this
without adding trusted authorities. Are they not incentive compatible? The
answer is no in the sense that each known solution only achieves convergence by
fixing at least one cost-imposing dimension of the environment so that it cannot respond
to strategic pressure. This is different from the challenge posed by Hurwicz,
who asked whether mechanisms could incentivize a stable enforcement function,
not assume one into existence beyond the game.

In games that converge through repeated play with Bayesian updating, achieving agreement
requires agents to update beliefs about a fixed state space with common
priors and stationary likelihoods (Milgrom \& Roberts; Fudenberg \& Levine).\cite{MilgromRoberts1990,FudenbergLevine1998} 
Agreement can also be induced through public randomness or correlated devices, but
such approaches work where randomness is unintelligible and thus
unactionable as in Myerson’s \emph{Great War} paper or Aumann’s correlated equilibrium.\cite{MyersonGreatWar,Aumann1974}  
Deviation in these approaches is only costly once incorrect beliefs can induce
suboptimal outcomes, so coordination unravels once agents can influence signal
generation, timing, or interpretation.

This same problem characterizes mechanisms that rely on fixed type
distributions, posted prices, or deterministic rules. Games that exploit
Byzantine fault tolerance, posted-price structures, or fair schedulers are only
stable when honesty, prices, and timing are strategically insulated: once those
dimensions are endogenized, security and convergence collapse. Approaches leveraging 
regret minimization and multiplicative weights exhibit the same structure:
they require stationary payoffs and non-strategic feedback, and cycling
reappears as soon as agent actions reshape allocations.

Across all cases, designers implement outcomes by asserting load-bearing structures
that resist strategic manipulation, not optimizing the enforcement
function itself through strategic play. Incentive compatibility requires
the latter, but this seems absurd: how can costs be both endogenously generated
and non-manipulable? Such opacity is self-evidently impossible in direct mechanisms,
where implementability demands truthful preference revelation and the need for
Maskin monotonicity further prevents agents from commingling multiple values into
non-scalar reports and sharing their preferences obliquely. Any solutions must
therefore lie outside the class of direct mechanisms or their
revelation-equivalent indirect counterparts.

As we shall see, the techniques which create endogenous unactionability do in fact 
block reduction under the Revelation Principle. But discussing how indirect mechanisms
can preserve opacity requires a way to reason about how mechanisms can protect
informational privacy. To provide these, our next section introduces the concepts of
\emph{Myerson} and \emph{Non-Myerson Layers} to show how constructs composed of
the latter can sustain the kinds of compartmentalization which block agents
from verifying the true preferences of other agents in the mechanism, even
while aggregating their underlying preferences into a shared enforcement cost.

\section{Myerson and Non-Myerson Layers}

The Revelation Principle is a brilliant analytical reduction which demonstrates
that many indirect mechanisms can be reduced to simplified games in which agents
report types, the mechanism selects an outcome, and transfers enforce truthful
revelation. Within its domain, the principle provides extraordinary clarity,
allowing designers to reason about incentives without worrying about the structural
compartmentalization or passage of time in a game.

Yet many non-revelation-equivalent games exist which cannot be reduced to direct
mechanisms without breaking incentive compatibility. The techniques that block
reduction are fairly easy to implement technically, including strategic routing
(Renou \& Tomala), some probabilistic lotteries (Strack \& Mora), and asymmetric
information disclosure (Attar). Claims of universal reduction only hold under 
restrictive assumptions that exclude these phenomena.

These non\-revelation\-equivalent strategies become meaningful when mechanisms are
treated as \emph{constructs} composed of multiple layers. In such designs, each 
layer can serve a different function, yet condition on inputs from previous
layers and create outputs for subsequent ones. A boundary between layers exists
at any point where inputs are strategically indistinguishable from randomness
when analyzed in isolation.

For clarity, we call a layer a \emph{Myerson Layer} if its incentive structure can
be faithfully represented by a reduced game that acts on direct-type reports and
has verifiable outputs. These layers follow the same rules as direct mechanisms
and, when chained together in sequence, may be analyzed as if occurring
simultaneously even though it may requires time to conclude.

In contrast, a \emph{Non-Myerson Layer} has the properties of non-revelation-
equivalent indirect mechanisms, relying on forms of action within the mechanism
that do not admit reduction without loss of essential structure, and producing 
outputs that encode preferences obliquely.

The distinction matters with circular mechanisms, as any sequence of Myerson Layers
collapses into a single Myerson Layer under the Revelation Principle, but
constructs with Non-Myerson Layers are irreducible because the informational
structure of at least one layer blocks reduction to direct type reports.

\begin{figure}[t]
\begin{narrowfigure}
    \includegraphics[width=\linewidth]{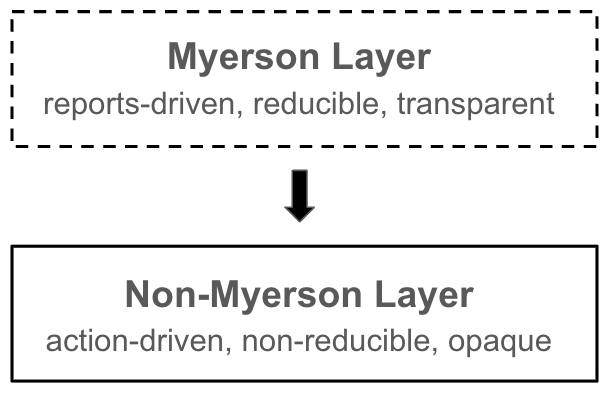}
  \caption{
  \textbf{Myerson Layer preceding Non-Myerson Layer.}
  The Non-Myerson layer acts on transparent inputs projected into it from the previous layer, and is able to reconstruct the intent behind the projected preferences.
  }
  \label{fig:myerson-nonmyerson}
  \end{narrowfigure}
\end{figure}

This makes \emph{constructs} useful conceptual tools for conceptualizing where
the demand for credibility and supply of enforcement happen in circular mechanisms. In classical
designs reliant on external authorities, enforcement is typically provided by
upstream institutions that project outputs downstream where they appears as 
exogenously-fixed honest types or other constraints. This is shown in Figure~\ref{fig:myerson-nonmyerson}, which shows 
a \emph{Non-Myerson Layer} preceding a \emph{Myerson Layer}.

\begin{figure}[t]
\begin{narrowfigure}
    \includegraphics[width=\linewidth]{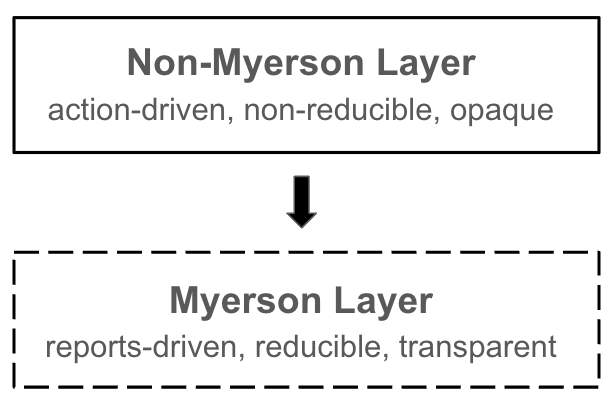}
    \caption{
  \textbf{Non-Myerson Layer preceding Myerson Layer.}
  The Myerson layer conditions on preferences expressed in the preceding layer which appear projected into it in forms indistinguishable from randomness.
  }
  \label{fig:nonmyerson-myerson}
  \end{narrowfigure}
\end{figure}

Such arrangements can affect the viability of trade-offs in downstream layers.
A canonical example is budget balance in the Myerson-Satterthwaite setting,
where adding a trusted auctioneer implicitly assumes an upstream layer to
incentivize its provision. When this layer is removed in revelation-equivalent
constructs, budget balance fails downstream as the mechanism must internalize 
the cost of generating enforcement that was previously incentivized externally.

Constructs in which Myerson Layers project information into Non-Myerson Layers
are rarer, but can be useful when the second layer introduces opacity, delay, or strategic
aggregation to the preferences revealed in the first, transforming the original
inputs into signals that emerge staggered in time. An example is sequential
voting systems with portfolio assembly, where individual votes cast in the first
layer may not be immediately visible in the outputs of the second. Blockchains
with private mempools show how this design can add obfuscation and delay which
frustrates attempts to reconstruct whose preferences are reflected in aggregated
outcomes at any point in time. This is depicted in Figure~\ref{fig:nonmyerson-myerson}, which shows a \emph{Myerson Layer} preceding a \emph{Non-Myerson Layer}

\begin{figure}[t]
\begin{narrowfigure}
    \includegraphics[width=\linewidth]{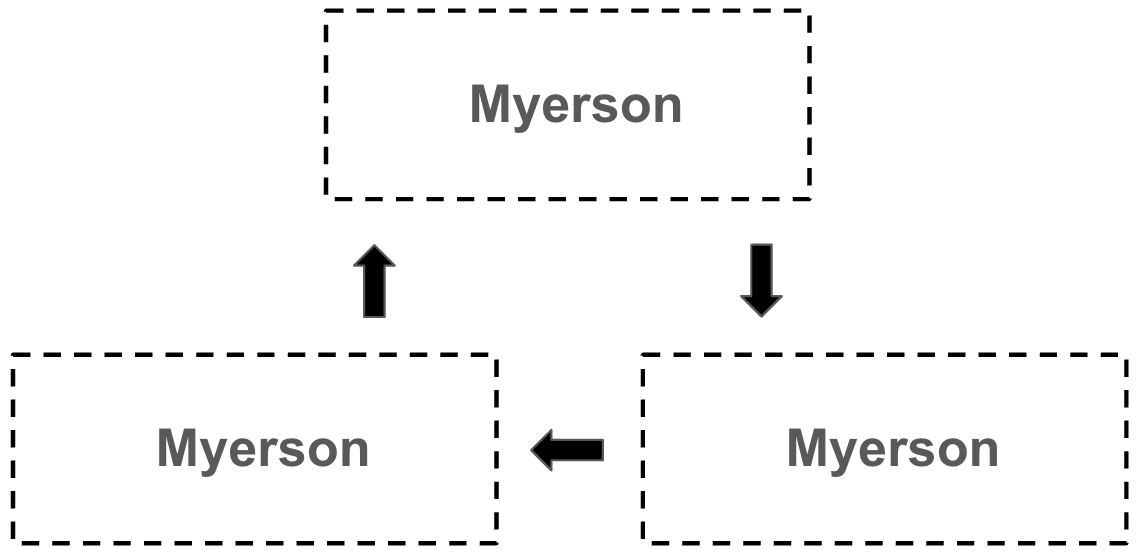}
  \caption{
  \textbf{Direct Mechanism as a Construct of Myerson Layers.}
  Multiple Myerson Layers chained into a circular construct form a Penrose-style mechanism which is locally coherent but globally incoherent.
  }
  \label{fig:myerson-construct}
  \end{narrowfigure}
\end{figure}

These differences become decisive when layers are chained into circular
constructs. Consider a construct consisting solely of Myerson Layers, which the
Revelation Principle collapses into a single layer in which outputs from any
round are transparent inputs to the next, as shown in Figure~\ref{fig:myerson-construct}.

\begin{figure}[t]
\begin{narrowfigure}
    \includegraphics[width=\linewidth]{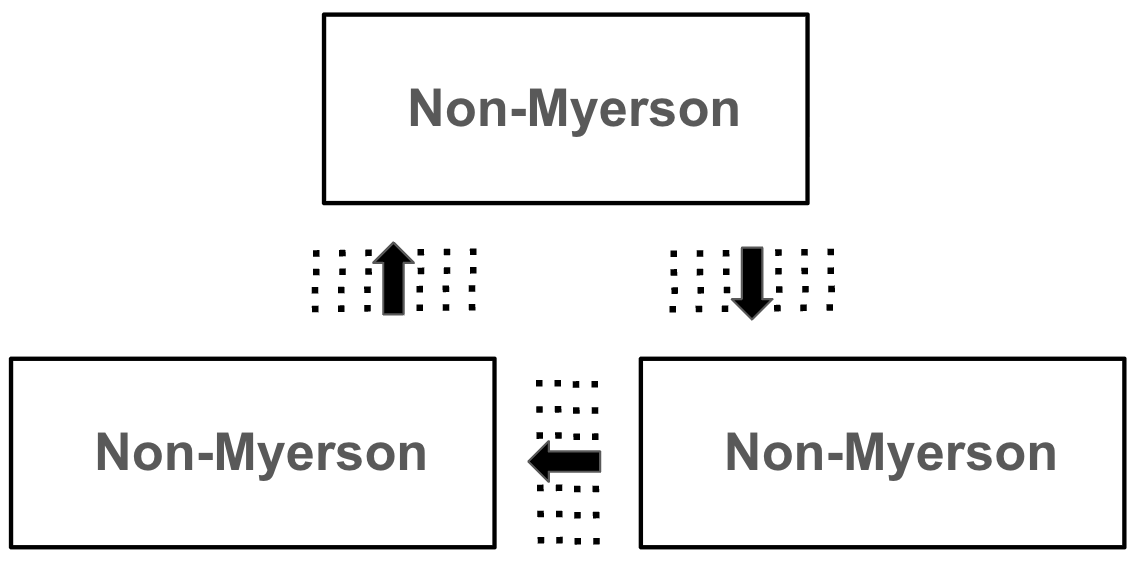}
  \caption{
  \textbf{Indirect Mechanism as a Construct of Non-Myerson Layers.}
  Multiple Non-Myerson Layers chained into circular constructs lose verifiability as privacy walls form between layers.
  }
  \label{fig:nonmyerson-construct}
\end{narrowfigure}
\end{figure}

Circular mechanisms with multiple Non-Myerson Layers are needed to preserve
forms of informational compartmentalization. While each layer still acts on
information projected into it from previous layers, its inputs need not be
reducible to direct type reports and may include the non-scalar outputs of 
previous layers that preserve ambiguity of intent. This is shown in Figure~\ref{fig:nonmyerson-construct}.

Across the lossy informational boundaries that separate layers in these
constructs, which we call \emph{privacy walls}, agents can observe actions
without reconstructing intent. They must, in effect, peer through a glass
darkly. In the next section, we show how these privacy walls combine with
non-scalar values to enable a new type of strategy that re-opens the door 
to incentive compatibility in conditions of decentralization.

\section{From Privacy Walls to Selective Disclosure}

Because direct mechanisms are evaluated in a static frame, the Revelation
Principle forces its circular mechanisms to operate
sequentially, as anything else implicitly breaks reduction (Myerson, 1986). 
Multiple layers may operate in parallel in non-revelation-equivalent mechanisms, however, 
permitting \emph{selective disclosure strategies} to emerge, where agents pierce 
privacy walls and reveal preferences directly to other participants.

\begin{figure}[t]
\begin{narrowfigure}
    \includegraphics[width=\linewidth]{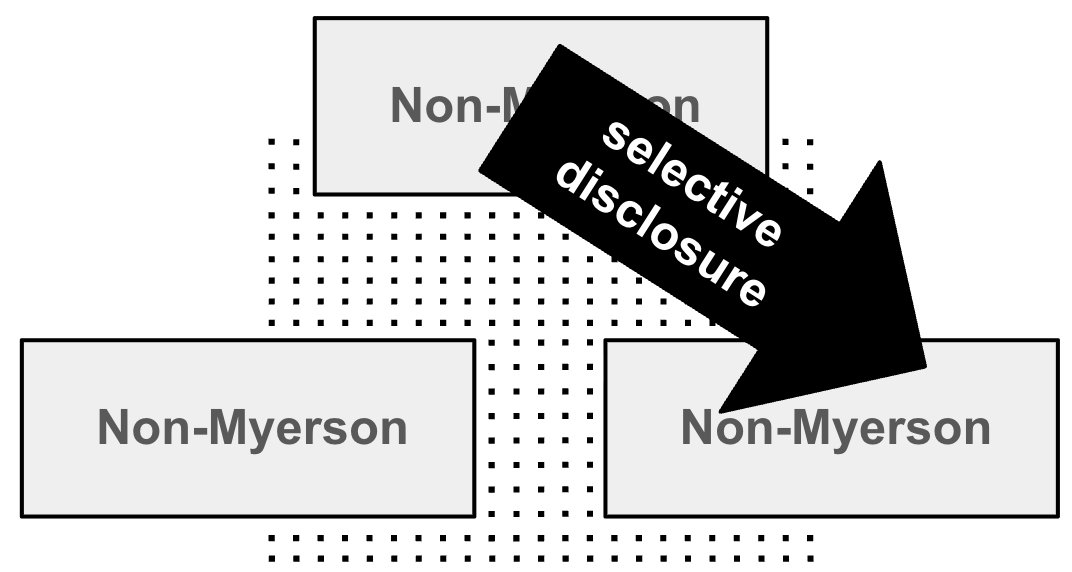}
  \caption{
  \textbf{Selective Disclosure as Communications Strategy.} 
  Agents pierce privacy walls and communicate preferences directly through Non-Myerson meta-games the mechanism encourages to be created in private channels where speech is not tracked or penalized, but whose output is a structured input returned to the mechanism which drives the mechanism into equilibrium. This makes it a part of the underlying game, not the "null case" of player "exit" or "opt-out" as retreats to non-enforcement environments are traditionally modelled in implementation theory.}
  \label{fig:selective-disclosure}
\end{narrowfigure}
\end{figure}

Such disclosures exist in many auction designs. They are common in
\emph{off-book trading mechanisms}, where players must choose between submitting
anonymous limit orders to a central book or negotiating privately with dealers
``upstairs''. They also appear in FCC-style spectrum auctions, where players may
tacitly disclose geographic priorities. In blockchains, similar disclosures
occur whenever users sell transactions to miners in return for privately
negotiated benefits.

As in other mechanisms, strategies involving the disclosure of preference maps
create a two-sided risk. Agents who offer sincere interpretations expose
themselves by making their internal structure legible, while agents who receive
disclosures face the risk of strategic manipulation. A blockchain user who sells 
transactions to a miner may be signalling interest in discounted inclusion, but 
their true objective may be to lower overall security and make deviations cheaper 
within the mechanism. An informed counterparty may rationally decline such a trade.

Because the act of piercing a privacy wall connects otherwise compartmentalized 
information spaces, and the base-layer mechanism only penalizes deviations involving 
non-scalar outputs from those compartments, selective disclosure strategies create new 
meta-games outside the mechanism’s core enforcement regime. Players have a strategic 
choice of which regime to use for preference revelation, and the economic consequences 
of deception can vary based on which communication channel is chosen.

The non-scalar messages that protect privacy on the base layer then reinforce the 
informational integrity of the meta-game, as players engaging in strategic disclosure 
cannot provide verifiable evidence of consistent beliefs with their preferences as 
revealed obliquely under the main enforcement regime. Where the mechanism makes playing 
such games rational for participants, it essentially induces kenosis: pulling away from 
players to create a safe space for them to interact that it cannot observe, but that 
will let it intuit, insulated from its influence, the marginal utility of its own 
enforcement function.

To see the solution more clearly, consider a case where Alice asks Bob to take an
action that lowers security in the mechanism. From Bob's perspective, Alice may 
be attempting to convert a global public good (security) into a dyadic trust good, 
capturing the savings while Bob absorbs the risk. Alternatively, her offer may be 
sincere, with Alice offering Bob benefits for helping her lower in-mechanism security. 

And while some may argue that Bob lacks all rational grounds to judge Alice’s 
proposal, we cannot restrict Bob deciding himself on the basis of his private 
preferences. The mechanism may not be able to observe those directly, but Bob 
is the decision-maker who determines their relevance. And there is no harm to 
letting him do so, for if Bob already prefers lower security, he needs no persuasion. 
Alice’s suggestion only becomes decisive in the exact case when Bob’s strategy 
changes based on it. And in that case her lack of credibility justifies his 
maintaining a high-security equilibrium, while his trust in her rationalizes 
lowering it.

From the perspective of the mechanism, Alice is opening a Non-Myerson Layer whose 
output and existence is opaque to the base-layer, operates under a different enforcement
regime and results in a non-scalar output that is indistinguishable from one produced
under the base-layer, and will invisibly commingle Bob's other preferences with his 
assessment of her credibility. The mechanism can see Bob’s choice without knowing what 
preferences motivates it. But it does not need this information: if Bob trusts Alice 
when its enforcement regime could protect him from loss, its level of protection is excessive.

The mechanism is essentially inducing a counterfactual game it cannot observe, 
one which nonetheless signals the marginal utility of its own enforcement function back 
to it in a form indistinguishable from normal gameplay. And this signal is credible precisely 
because Bob cannot know which activities on the base-layer network represent Alice, so 
no representations she can make do not invite questions of credibility.

For readers accustomed to direct mechanisms, it may seem counterintuitive that a
mechanism never observes the preferences that are decisive. But it does not need
to. When enforcement levels are above equilibrium, private communications are 
cost-saving and attractive. As they converge toward equilibrium, the risks 
of disclosure grow while its benefits shrink, and disclosure strategies become 
less attractive on the margin. The mechanism essentially incentivizes retreat 
to alternate enforcement regimes in the proportion needed to drive its 
marginal utility of enforcement into optimal equilibrium.

Yet these selective disclosure strategies exist only in non-revelation-equivalent
mechanisms. In the next section we show this formally, demonstrating that no
direct report Alice can offer Bob avoids collapsing into epistemic claims whose
credibility cannot be established within the mechanism itself.

\section{Epistemic Uncertainty from Unrepresentable Values}

Throughout this paper, we use \emph{uncertainty} in the sense articulated by
Knight, Keynes, and G.~L.~S.~Shackle: not as a variable that can be calculated or
approximated probabilistically, but as \emph{unknowability} -- the absence of
any representation that can be rationalized within an internally coherent
evaluative framework. As Keynes famously wrote:

\begin{quote}\small
The sense in which I am using the term is that in which the prospect of a European war is uncertain\ldots{} or the price of copper and the rate of interest twenty years hence\ldots{} About these matters there is no scientific basis on which to form any calculable probability whatever.\footnote{Keynes, J. M. (1936). \emph{The General Theory of Employment, Interest, and Money}, Chapter~12. Macmillan.}
\end{quote}

For the sake of convenience, we refer to whatever variable motivates agents to
accept or reject selective disclosure proposals as \emph{trust}. This is not meant as an 
ontological statement. We do not attribute moral qualities to trust, nor
deny that it may be the product of utilitarian calculation. We merely name a
variable that motivates action when enforcement levels are decisive at
the margins. So we state that an agent acts on trust whenever they accept that statements offered
by a counterparty may be false, yet prefer to act as if they are true. Agents
who withhold trust prefer to assume statements may be false and instead rely on
the baseline guarantees enforced by the mechanism.

This approach contrasts with much modern work in computer science, which seeks
to eliminate uncertainty by expanding the state space through additional
parameters, beliefs, or utility dimensions. We do not dispute the usefulness of
such methods, but deny that they can achieve endogenous unactionability.

The reason is that regardless of how one conceptualizes trust -- whether as a
psychological variable or a non-scalar value that can be formally modelled -- 
representing it as a direct type to the base-layer collapses the information it is
meant to convey. The problem is connected to the way security and trust become
substitutes across uncorrelated games when agents have the option of lowering
enforcement costs in one by extending trust in another. We can show this formally.

\subsection{Trust as an Unparameterizable Domain Space}

In informationally decentralized systems such as consensus mechanisms, 
deviation must be discouraged without recourse to external enforcement. 
Generating security is therefore costly. Let $S \ge 0$ denote the level 
of security imposed by the mechanism, with cost function $K(S)$, where

\[
K'(S) > 0.
\]

Let $\tau$ denote the \emph{trust environment}: an aggregate property capturing
how willing participants are to refrain from deviation absent strong
enforcement. Let $R(S,\tau)$ denote residual cheating risk, where

\[
\frac{\partial R}{\partial S} < 0,
\quad \text{and} \quad
\left| \frac{\partial R}{\partial S} \right|
\text{ decreases as } \tau \text{ increases}.
\]

Trust and security are substitutes in this environment: trust reduces the
marginal utility of additional enforcement. When Alice proposes that Bob
cooperate with her to reduce security, she is implicitly suggesting that welfare
can be jointly increased by reallocating resources from enforcement to other
forms of utility at the margin.

Welfare improvements require changing the socially optimal level of security,
which solves

\[
\min_{S \ge 0} \; K(S) + R(S,\tau),
\]
with first-order condition
\[
K'(S^*) = -\frac{\partial R(S^*,\tau)}{\partial S}.
\]

It follows immediately that the optimal security level $S^*(\tau)$ is strictly
decreasing in $\tau$.

\subsection{The Trust Parameter Cannot Be Elicited}

To compute $S^*(\tau)$, the mechanism must know $\tau$. In systems with external
authorities, this information may be inferred institutionally. In
informationally decentralized mechanisms, however, the mechanism would need to
elicit $\tau$ as a type.

This is impossible.

A declaration of the form ``my trust level is $\tau$'' cannot distinguish a
genuinely high-trust agent from a low-trust agent strategically claiming higher
trust in order to reduce penalties. The type and its anti-type generate identical
messages.

\subsection{Gödel-Type Collapse}

While any informationally decentralized mechanism that must optimize its own
enforcement levels requires $\tau$, any attempt to represent it directly is
strategically manipulable in exactly the dimension it is meant to measure.
Encoding trust destroys its informational role. This is a Gödel-type phenomenon:
certain economically necessary variables are unrepresentable within mechanisms
because any attempt to encode them renders the encoding meaningless.

This collapse creates \emph{fundamental unrepresentability}. The mechanism requires
$\tau$ to compute $S^*(\tau)$, but $\tau$ cannot be known ex ante, inferred from
reports, or represented within the mechanism without contradiction. This is not
epistemic ignorance resolvable through learning; it is structural. The mechanism
cannot observe the counterfactual worlds in which different trust levels would
have prevailed, nor price security optimally without exposing itself to over- or
under-enforcement.

At first glance, this conclusion may seem too strong. One might object that the
determinants of credibility may be discovered and formalized. And if credibility 
is eventually discovered to be a well-defined variable, why can it not be elicited 
by the base layer directly?

Except our challenge is not whether credibility can be defined, but whether it can 
be made \emph{actionable} within the base layer without breaking incentive
compatibility. And while a mechanism could theoretically condition enforcement on well-defined 
types, it would need to filter its reports to privilege credible agents. And this 
would exclude the preferences of non-credible agents from aggregation, driving 
enforcement costs away from social optimal levels. Incentive compatibility requires 
optimizing over all preferences, not merely a subset of trusted players.

And if correlation is possible despite a lack of semantic definition? In such a case 
there apparently exists a positive, exploitable correlation that motivates selective disclosure 
strategies. In that case, misrepresentation is profitable at the margin, and agents 
who extend trust are systematically exploited until they update their beliefs defensively 
until the correlation no longer supports profitable inference. And in equilibrium, the 
only stable such point is when the correlation is zero at the margin.

\subsection{Uncertainty and Indirect Mechanisms}

Indirect circular mechanisms are the only class capable of resolving this
problem, because they allow participants to play games in parallel under different
enforcement regimes and treat the outcome as a counterfactual judgment on the
marginal utility of enforcement within the mechanism.

Within each game, the other appears as a Non-Myerson Layer whose output is opaque
and subject to misinterpretation. Each nevertheless serves as a venue in which
rational judgment may expressed. And while the preferences and cognitive frameworks
agents use may differ across layers, actions remain credible across them, as 
representations in the parallel game are penalized on the base-layer.

In this way, indirect mechanisms convert an unrepresentable parameter into an
observable gradient without collapsing uncertainty into a reportable type. The
solution is unimplementable without a base-layer game consisting of Non-Myerson 
Layers that generate privacy walls, creating potential for agents to bypass them, and letting the mechanism treat the 
result as an opaque referendum on the efficiency of its marginal allocation of resources 
to the enforcement regime of the mechanism.

It follows that any informationally decentralized mechanism that must optimize
its own enforcement levels cannot be direct. What it must observe is a
Gödel-type object: fleeting, disguised, mixed invisibility into a non-scalar message
and appearing only when decisive. And structurally impossible to represent directly, 
despite being capable of incentivization and subject to rational consideration.

\section{From Uncertainty to Incentive Compatibility}

This paper began with a narrow problem: how incentive compatibility can be
sustained in informationally decentralized environments. In the preceding
sections we showed that the solution lies in the endogenous generation of costs
that agents cannot neutralize within a mechanism, and that this is possible in
action-based mechanisms with non-scalar messages and privacy walls.

Selective disclosure plays a decisive role in this process and yields its simple
equilibrium logic. When enforcement is too strong, agents have incentives to
cooperate to reduce unnecessary security costs. When enforcement is too weak,
agents retreat to trust-free strategies and demand stronger guarantees,
preferring strategies that provide greater protection against revision. Once
enforcement is correctly calibrated, neither revealing nor concealing private
structure can improve expected welfare, selective disclosure ceases to be
profitable, and security remains stable in equilibrium.

Seen in this light, uncertainty is not an epiphenomenon but the functional
substitute for external authority in decentralized systems. And incentive
compatibility can emerge more generally as self-stabilizing security levels now
provide a common \emph{num\'eraire} against which other forms of utility can be
expressed. Appendix~D shows a practical example of a mechanism exhibiting this
dynamic.

The technique may only be implementable within a narrow subclass of indirect
mechanisms, but it exists. In the next section, we revisit the large body of
impossibility results in computer science and economics which claim it cannot,
showing that their conclusions exclude indirect mechanisms by assumption, often
for reasons their authors do not recognize can be relaxed.

\section{Explaining Impossibility}

Across economics and computer science, impossibility results share a common
premise: that all strategic interaction occurs under a single enforcement
regime. The edifice of mechanism design is typically formulated under this assumption: removing the possibility that agents might rationally shift
coordination into lower-security domains created by the mechanism itself.

This possibility lay outside the scope of early theorists. Because Hurwicz
examined allocation in environments whose setting was practically
indistinguishable from the state of nature, he did not model the case in which successful 
implementation might create an enforcement regime from which strategic retreat 
might be possible. His impossibility result is therefore
correct for the class of mechanisms he analyzes, but does not extend to
mechanisms that give agents the freedom to coordinate under reduced
enforcement guarantees, and incentivizes them to do so for its own purposes.

Other impossibility results rest on the same assumption. Arrow’s impossibility
theorem, the Gibbard--Satterthwaite result, and Maskin’s monotonicity condition
all presuppose that preferences must be represented as semantically meaningful
scalar reports, that communication is verifiable, and that deviation is
disciplined by a symmetry of enforcement costs. These results do not assert that
incentive compatibility is impossible in informational decentralization per se,
but they encourage the conclusion that strategic interaction must occur under a
single enforcement regime.

Canonical results from computer science reinforce the same assumption. Developed
in foundational papers by Lamport, Shostak, and Pease, and later formalized by
Bracha and Toueg, results on the limits of adversarial tolerance assume that
algorithms behave according to fixed rules and execute changes as \emph{atomic
steps}.\cite{LamportShostakPease1982,BrachaToueg1985,FLP1985}
 These models are defined to treat actions as costless, abstract away Knightian uncertainty, and assume deviation cannot be punished through exposure accumulated
over time. Later results such as the FLP theorem operate in this same framework,
in which non-termination and inconsistency arise because no mechanism can
impose asymmetrical costs that diverge over time.

In more recent work, including the so-called TFM literature, we find further
arguments discouraging attention to indirect mechanisms as a viable solution
class, with some authors even asserting analysis can be entirely restricted to
the study of direct mechanisms. Many of these results fail to generalize due to 
improper reduction under the Revelation Principle, often a result of changing a 
two-sided, multi-parameter and dynamic game into a one-sided, single-parameter,
static auction that cannot even in theory implement the same set of outcomes. A
contributing source of difficulty is that off-chain coordination in TFMs
naturally invites action under multiple enforcement regimes, which is
interpreted to suggest the presence of multiple direct mechanisms rather than
circular constructs with interacting non-revelation-equivalent layers.

Related work inherits these restrictions implicitly through modeling choices.
Bayesian analyses assume that learning takes place over static distributions
within a single enforcement framework. Posted price models presuppose endogenous
unactionability under highly restrictive conditions. And the pervasive
assumption that miners ``faithfully implement'' protocols is repeatedly used to
avoid eliciting preferences from them, an omission that leads to recurring
problems with interdependent valuations once miners re-enter the game as strategic
players. These results generate impossibility not because their incentives are
ill-posed, but because their frameworks were only intended to handle direct
mechanisms, not indirect constructs with multiple interacting
Non-Myerson Layers.

Philosophically, the category error that emerges is the inverse of that
identified by Gilbert Ryle in \emph{The Concept of Mind} (1949). In that book,
which popularized the phrase ``ghost in the machine,'' Ryle critiqued Descartes
for assuming the existence of an immaterial spirit simply because the mind seems
to have agency. The error in our time is typically reversed: analysts retreat to
deterministic models in part because networks are made of machines. Impossibility
reappears quietly because deterministic execution cannot accommodate semantic
ambiguity.

This same pattern appears in economic papers on the limits of permissionless
consensus, where economists import constraints from computer science as fixed
technical limits and then restrict possibility within those bounds. Some papers
assume verifiability is needed for incentive compatibility when that is not the
case, as with Groves-Ledyard or dynamic pivot mechanisms. Others argue that
consensus mechanisms must be voting systems, treating verifiability and
voting-like behavior as necessary primitives and excluding stabilizing forms of
off-chain coordination by assumption.

But this suggests a puzzle: if indirect mechanisms can in principle
escape classical impossibility results, why has the literature repeatedly failed
to converge on them? Why do attempts to relax these assumptions so often conclude
that nothing fundamental changes?

\section{The Gravity Well of Direct Mechanisms}

The answer seems to be that mechanisms governed by the Revelation Principle exert a
strong cognitive pull. Once a model satisfies even a subset of its defining
assumptions—costless communication, verifiable reports, fully specified state
spaces—it is drawn back into what we call the \emph{gravity well} of revelation,
where mechanisms collapse into direct forms unless costs are imposed under
uncertainty.
This pattern appears repeatedly across the literature in a series of near-miss
attempts to escape impossibility by relaxing dimensions of the standard
framework one at a time.

\textbf{Time:} a large body of work introduces dynamics --- rounds, epochs,
repetition, or sequential interactions --- yet preserves transparency,
verifiability, and report-based reasoning. In these models, time functions as an
index over observations rather than as a generator of uncertainty. Learning
sharpens inference, beliefs converge, and temporal structure accelerates the
collapse back into revelation-equivalent paralysis.

\textbf{Collusion:} many models enrich the strategy space by allowing agents to
collude, coordinate, or form coalitions. This introduces high-dimensional
preferences, but these cannot be used to implement social choice rules, as their
existence generates productive ambiguity about agent motives and preferences for
future payoffs. As a result, stabilizing forms of cooperation that increase
welfare are removed alongside suboptimal collusion, and off-chain agreements are
recast as behaviors to be detected, punished, and eliminated.

\textbf{Risk:} a prominent line of work attempts to impose costs by making attacks
expensive through fees, slashing, or resource expenditure. When such costs can be
priced ex ante, however, they enter agents’ decision-making frameworks as risks
over known states rather than as exposure to uncertainty over time. Deterrence
then fails, as it is modeled through equilibrium selection rather than through
forced exposure to front-loaded costs under Knightian uncertainty.

\textbf{Permission:} many proposals reintroduce trusted committees, governance
layers, or validator sets. These moves can succeed precisely because they smuggle
in additional Myerson or Non-Myerson Layers that live outside the layer under
analysis. Yet because these changes are framed as shifts in trust assumptions
rather than informational structure, they are treated as orthogonal to the
incentive problem instead of as alterations that will also collapse once subject
to systemic rationality.

\textbf{Stabilizers:} posted prices, fixed fee schedules, and deterministic
inclusion rules appear to move away from the Revelation Principle by constraining
behavior in other ways. In practice, they deepen the gravity well by making
actions legible, deviations interpretable, and learning faster. They are steps
toward direct mechanisms, not away from them.

\textbf{Authority:} the introduction of stakers, validators, and internal
governance bodies shifts authority from the environment into the mechanism. Once
authority becomes endogenous, however, control itself becomes a source of
utility, and the mechanism is tasked with incentivizing its own enforcement,
reintroducing the circular paradox diagnosed by impossibility results.

In retrospect, there is no smooth path from the direct world to the indirect one.
The transition appears to require a discontinuous shift in perspective across
multiple dimensions: reports versus actions, static versus dynamic mechanisms,
scalar versus non-scalar messages, risk versus uncertainty, monotonicity versus
non-monotonicity, verifiability versus privacy, and ex ante costs versus ex post
benefits.

From the perspective of the direct mechanism, this leap is invisible. It requires
abandoning too much analytic scaffolding to even conclude that the solutions are
tractable. Approaches that rely on privacy appear incoherent in settings that
demand verifiability, and some variables can only be made decisive by rendering
them semantically meaningless. Matters are further complicated by the possibility
that multiple cognitive models may coexist, pulling the concept of utility itself
into question.

As such, classical impossibility results are neither surprising nor discredited.
They remain faithful descriptions of what occurs inside the gravity well of the
Revelation Principle. From the perspective of the indirect mechanism, however, the
direct mechanism emerges as the degenerate case in which impossibility arises
because uncertainty is resolved immediately, a single enforcement regime is
imposed by fiat, and values that matter become semantically impossible to express.

\section{Conclusion}

This paper began from a narrow question about incentive compatibility in
distributed consensus and arrived at a broader reconsideration of how
cooperation can be sustained without external authorities, showing the
conditions under which incentive compatibility is possible.

Seen from this perspective, the power and the limits of the Revelation Principle
become clearer. The principle is extraordinarily effective within the realm it
was designed to govern: static or acyclic environments with costless speech and
fully specified state spaces. But in indirect circular mechanisms where multiple
enforcement regimes exist, improper reduction can strip away degrees of freedom
by obliterating strategic ambiguity. Much like a black hole that collapses
structural coherence at its event horizon, reduction transforms meaningful
information into semantic indecipherability. Nothing essential can be lost only
if nothing opaque mattered to begin with.

The broader implications are worth noting. In his \emph{General Theory} (1936),
Keynes extended Knightian uncertainty into macroeconomics, arguing that liquidity
traps persist not because agents miscalculate risk, but because stasis prevails
when uncertainty looms over the value of potential investments. On a more modest
level, what has been missing in mechanism design is a way out of this same trap:
a way for uncertainty to discipline behavior in ways that motivate price
discovery instead of punishing it. By forcing agents to reveal strategic
preferences for playing different games, indirect mechanisms can accomplish
this, transforming uncertainty into a cost that can be rationally borne, shared,
and priced without ever being fully resolved.

We believe that mechanism design must expand its vocabulary to incorporate these
techniques and concepts. In mechanisms with routing strategies, variable-time
hashing lotteries, and looping Non-Myerson constructs, sacrificing verifiability
is not a defect but the price paid to sustain the forms of informational privacy
that permit the mechanism to see its own value reflected indirectly in the
willingness of actors to risk loss in its absence.

The lesson returns us, in many ways, to our mythological stories of creation.
Where stabilizing costs can be paid outside a system, stability can follow.
Where they cannot, they must be borne within the system in a manner that renders
them external and incomprehensible to its own rational framework.
\clearpage


\newpage

\appendix

\section{Terminology}

This appendix collects and standardizes the key terms introduced in the paper.
Its purpose is not to introduce new concepts, but to fix language so that ideas
are referred to consistently and unambiguously throughout.

\textbf{Mechanism.} A mechanism is a process that specifies admissible actions,
information structure, and outcome rules governing interactions between agents.
Following Hurwicz, a mechanism maps actions or messages into outcomes.

\textbf{Incentive compatibility.} A mechanism is incentive compatible with
respect to a social choice rule if, for each agent, acting according to the
strategy prescribed by the mechanism is optimal given the strategies of others,
so that the induced equilibrium implements the intended social choice rule.

\textbf{Direct mechanism.} A direct mechanism is one in which agents report types
or preferences in a single interface, after which the mechanism selects an
outcome and transfers enforce truthful revelation. Direct mechanisms are the
canonical objects governed by the Revelation Principle.

\textbf{Indirect mechanism.} An indirect mechanism is one in which agents express
preferences through actions taken within the mechanism rather than through
direct type reports. Outcomes depend non-trivially on how actions are aggregated,
sequenced, and interpreted.

\textbf{Construct.} A construct is a mechanism composed of multiple internally
chained layers. Constructs, rather than individual layers, are the primary
objects of interest in circular and self-referential mechanisms.

\textbf{Layer.} A layer is a subcomponent of a construct in which agents condition
on inputs and project outputs into subsequent layers in ways that appear to be
unactionable when layers are studied in isolation.

\textbf{Myerson Layer.} A Myerson Layer is a layer whose incentive structure can
be faithfully represented as a direct mechanism acting on type reports and
projecting its output as publicly observable randomness or state changes. Any
sequential composition of Myerson Layers may be reduced to a single Myerson Layer
under the Revelation Principle.

\textbf{Non-Myerson Layer.} A Non-Myerson Layer is a layer whose incentive
structure is not revelation-equivalent. Such layers rely on action-based
incentives, ambiguous signals, dynamic preference aggregation, delayed
resolution, or strategic routing, and cannot be reduced without loss of
essential structure.

\textbf{Privacy wall.} A privacy wall is an informational boundary created by
Non-Myerson Layers across which actions and outcomes are observable, but
underlying motivations cannot be uniquely inferred. Privacy walls preserve
ambiguity of intent by permitting non-scalar variables to motivate action and
prevent agents from conditioning strategies outside the mechanism on fully
reconstructible causal explanations of behavior within it.

\textbf{Circular mechanism.} A circular mechanism is a mechanism whose outputs
re-enter itself as inputs in future stages of play, so that agents’ present
actions shape the future strategic environment in which they and others must
act. In circular mechanisms, learning and revision are endogenous, and incentive
compatibility depends on how the mechanism disciplines attempts to revisit or
unwind past decisions.

\textbf{Risk.} Risk refers to uncertainty over a fully specified and stable set
of possible states, where the mapping from actions to outcomes is known and
probabilities can be meaningfully assigned \emph{ex ante}. Under risk, agents can
evaluate strategies by computing expected utilities, price deviations in
advance, and revise behavior through belief updating without incurring
structural costs beyond those implied by known distributions.

\textbf{Knightian uncertainty.} Knightian uncertainty refers to uncertainty over
states, variables, or aggregation outcomes that are not defined at the time
action must be taken, and therefore cannot be assigned probabilities \emph{ex
ante}.

\textbf{Unactionability.} A constraint is unactionable if agents cannot condition
their behavior so as to offset, unwind, or neutralize its effects within the
mechanism, even when the constraint is observed or anticipated.

\textbf{Endogenous unactionability.} A constraint is endogenously unactionable if
it is generated by strategic interaction within the mechanism and, once
generated, cannot be neutralized, offset, or unwound by further strategic action
within that same mechanism.

\textbf{Informational decentralization.} A mechanism is informationally
decentralized if no agent or authority has access to the full state required to
determine outcomes or enforce behavior at the time actions are taken.
Information relevant to outcomes is dispersed across agents and realized only
through interaction over time.\newpage

\section{Illustrative Circular Mechanisms}

\begin{adjustwidth}{-0.75in}{-0.75in}
\begin{table*}[!t]
\centering
\small
\setlength{\tabcolsep}{5pt}
\renewcommand{\arraystretch}{1.05}
\begin{tabularx}{\linewidth}{
  p{3.4cm}
  p{1.6cm}
  p{2.4cm}
  X
}
\toprule
\textbf{Mechanism} &
\textbf{Type} &
\textbf{Revelation-Equivalent} &
\textbf{Location of Unactionability} \\
\midrule

VCG, Myerson auctions
& Direct
& Reducible
& Exogenous (verifiability, transfers, enforcement assumptions) \\

One-shot voting games
& Direct
& Reducible
& Exogenous (fixed rules, no revision, costless speech) \\

Repeated direct games (Bayesian updating)
& Direct
& Reducible
& Exogenous (fixed state space, common priors, stationary likelihoods) \\

Public randomization / correlating
& Both
& Reducible
& Exogenous (unmanipulable public randomness) \\

Posted-price mechanisms
& Indirect
& Reducible
& Exogenous (prices set outside the mechanism) \\

Continuous double auctions
& Indirect
& Irreducible (mostly)
& Mixed; exogenous via institutional rules, endogenous via market impact \\

Reputation systems
& Indirect
& Irreducible
& Endogenous (history-dependent payoffs, reputational effects) \\

Contract renegotiation mechanisms
& Indirect
& Irreducible
& Mixed; exogenous via courts, endogenous via renegotiation costs \\

Prediction markets (endogenous liquidity)
& Indirect
& Irreducible
& Endogenous (liquidity and price impact respond uncontrollably) \\

Proof-of-Work consensus
& Indirect
& Irreducible
& Endogenous (irreversible expenditure of hash) \\

Proof-of-Stake slashing
& Indirect
& Irreducible
& Mixed; Endogenous w/ Exogenous Governance \\

Saito Consensus
& Indirect
& Irreducible
& Endogenous (irreversible loss of continuation value) \\

\bottomrule
\end{tabularx}
\caption{A taxonomy of circular mechanisms with unactionability}
\end{table*}
\end{adjustwidth}
\clearpage


\section{Saito as an Implemented Instance of Endogenous Unactionability}

This appendix provides a constructive witness to the claims made in the paper. It
does not attempt to prove correctness, security, or equilibrium properties.
Rather, it shows how non-reducibility emerges in practice from the structural
requirements for incentive compatibility identified in theory.

\subsection{Layers, Aggregation, and Arbitrage}

Saito Consensus consists of three Non-Myerson Layers cycling in parallel. Each
layer aggregates preferences revealed by others into welfare-improving proposals
that adjust global prices. Three cross-layer properties are primitive structural
requirements for avoiding impossibility in theory:

\begin{itemize}
\item \textbf{Continuation value.} Benefits include options for continued play,
which are actively traded between participants as part of the aggregation
process, dividing transaction fees into mining and routing payouts.

\item \textbf{Costly uncertainty.} All strategies require participants to commit
to front-loaded and irreversible expenditures of utility that are fungible with
continuation value (fees, routing rewards, or resource consumption) under
uncertainty.

\item \textbf{Informational decentralization.} No trusted authorities exist in
any layer of the mechanism. Players act under uncertainty about the global
state, which is dispersed and aggregated only through play over time.
\end{itemize}

These properties apply to each of the three layers in the mechanism, which are
distinguished by their aggregation functions and the arbitrage opportunities
they expose.

\paragraph{User Layer}
\begin{itemize}
\item \emph{Aggregates:} Heterogeneous bundles of utility into transactions.
\item \emph{Arbitrages:} Differences between local execution opportunities and
global settlement prices, broadcast as transactions for processing and
settlement by other layers.
\end{itemize}

\paragraph{Routing Layer}
\begin{itemize}
\item \emph{Aggregates:} Bundles of transactions into provisional blocks.
\item \emph{Arbitrages:} Differences between marginal routing rewards and the cost
of providing ancillary collusion goods, producing efficient routing paths for
payout division.
\end{itemize}

\paragraph{Chain Resolution Layer}
\begin{itemize}
\item \emph{Aggregates:} Bundles of blocks into chains with a cost of revision.
\item \emph{Arbitrages:} Discrepancies between current aggregate security
preferences and expected future settlement conditions, broadcast as payouts
that simultaneously propose adjustments to global prices.
\end{itemize}

Each layer acts both as an aggregator of upstream preferences and as a generator
of downstream arbitrage opportunities. No layer determines final meaning in
isolation. Convergence is emergent, and the cost of revision increases with block
settlement depth in the game tree.

\subsection{Trust, Uncertainty, and Exposure}

Trust is a semantically meaningless variable. It is measured only indirectly,
through the shifts it induces from reliance on mechanism-enforced collective
settlement guarantees to coordination through private channels where
counterparty credibility is subject to time inconsistency.

These shifts take the form of selective disclosure strategies that pierce the
privacy walls of the mechanism. Where such strategies are rational, they improve
public welfare by optimizing the enforcement function. Where they are irrational,
they reduce the expected utility of the proposer.

\paragraph{User Layer}
\begin{itemize}
\item Selective disclosure of preference maps to routers.
\end{itemize}

\paragraph{Routing Layer}
\begin{itemize}
\item Selective disclosure of preference maps to users.
\item Selective disclosure of routing competitiveness to peers.
\end{itemize}

\paragraph{Chain Resolution Layer}
\begin{itemize}
\item Selective disclosure of willingness to bear future exposure.
\end{itemize}

All forms of selective disclosure become less rational as the marginal utility of
environmental trust converges toward equilibrium with the marginal utility of the
mechanism’s enforcement function.

\subsection{Non-Reducibility under the Revelation Principle}

The mechanism is a three-layer construct composed entirely of Non-Myerson
Layers, and cannot be reduced to a revelation-equivalent mechanism. Messages are
shared through public channels only as non-scalar values. Multiple strategies
block reduction.

\paragraph{User Layer}
\begin{itemize}
\item Routing strategies (Renou \& Tomala).
\item Selective disclosure (Attar).
\end{itemize}

\paragraph{Routing Layer}
\begin{itemize}
\item Routing strategies (Renou \& Tomala).
\item Selective disclosure (Attar).
\end{itemize}

\paragraph{Chain Resolution Layer}
\begin{itemize}
\item Routing strategies (Renou \& Tomala).
\item Selective disclosure (Attar).
\item Time-uncertain fixed-difficulty hashing lottery (Strack \& Mora).
\end{itemize}

Across layers, selective disclosure introduces additional Non-Myerson Layers
that observe mechanism outputs and produce non-scalar outputs reflecting the
attractiveness of coordination outside the mechanism’s enforcement guarantees.

No portion of the mechanism is reducible to a direct mechanism:
\begin{itemize}
\item No layer relies on truthful type reports.
\item All layers operate concurrently rather than sequentially.
\item Information flows are non-stationary and non-topologizable.
\item Actions project ambiguous signals rather than verifiable messages.
\item Some strategies skip layers through uncertain execution order.
\item Learning cannot model semantically undefined variables.
\end{itemize}

Attempts to analyze any single layer in isolation assume away the exogenous
structure provided by the construct as a whole, reintroducing verifiability only
by destroying the opacity that allows the mechanism to observe rational
responses to the marginal utility of enforcement.

Because all forms of value are fungible with continuation value, higher-
dimensional optimization emerges over message spaces containing utilities that
can be exchanged for continuation value. These trades aggregate into global
prices, inducing convergence.

\subsection{What This Appendix Does Not Claim}

This appendix does not claim:
\begin{itemize}
\item that Saito is unique;
\item that similar mechanisms are easy to design; or
\item that the theory provides a general construction algorithm.
\end{itemize}

It makes a narrower and stronger observation: \emph{a mechanism satisfying the
theoretical constraints identified through implementation theory does exist},
and when other incentive-compatible mechanisms are discovered, they will exhibit
the same structural properties.
\clearpage


\section{Interpretive Constraints and Irreducible Structure}

This paper deliberately avoids equilibrium characterizations and toy models that
represent trust, credibility, or willingness to bear exposure as actionable parameters. 
Any such formalization would reenact the collapse it diagnoses: decisive variables 
would become conditionable within the meta-theory, reintroducing the gravity well of 
revelation-equivalent mechanisms that these constructions are designed to escape. The 
argument therefore proceeds through structural contradiction, historical diagnosis of 
impossibility results, and constructive witness rather than internal equilibrium proofs. 
This is a methodological boundary, not an absence of rigor.

\subsection{Irreducibility and Forbidden Reductions}

The mechanisms studied in this paper are intentionally
\emph{non-revelation-equivalent}. Any analytical move that presupposes the
applicability of the Revelation Principle --- including reduction to direct
mechanisms, representation of decisive variables as scalar types, or evaluation
under a single enforcement regime --- alters the object under study and
invalidates conclusions drawn from it.

In particular:
\begin{itemize}
\item No agent possesses a type that encodes trust, credibility, or willingness to
bear exposure.
\item No report or message can truthfully represent preferences over enforcement
levels.
\item No single enforcement regime governs all strategic interaction.
\end{itemize}

Analyses that assume otherwise implicitly reintroduce exogenous authority or
verifiability and therefore collapse the mechanism into a different class.

\subsection{Non-Scalar Variables and Epistemic Boundaries}

Several variables that are decisive for equilibrium selection in these mechanisms
are \emph{non-scalar} and intentionally unrepresentable. These include trust,
credibility, and continuation value under uncertainty. Their role is not to be
measured directly, but to motivate action in ways that cannot be neutralized or
arbitraged within the mechanism.

The paper’s central claim is not that such variables are ineffable or
psychological, but that any attempt to encode them as types destroys the
information they convey. This creates epistemic boundaries --- privacy walls ---
across which actions are observable but intent is not reconstructible. These
boundaries are structural, not accidental.

\subsection{Parallel Enforcement Regimes}

Selective disclosure strategies operate by shifting coordination into
\emph{parallel games} governed by different enforcement regimes. These regimes
are not deviations, collusion, or exits from the mechanism, but are induced by its
design and feed back into it as structured but opaque signals.

The mechanism does not observe the content of these interactions. It observes
only the willingness of agents to bear exposure outside its guarantees. This is
sufficient to infer the marginal utility of its own enforcement function.

Any interpretation that treats off-mechanism coordination as exogenous or illicit
misses this feedback loop and misidentifies the equilibrium logic.

\subsection{Shiva, Entropy, and Continuation Value}

The role played by \emph{Shiva} in this paper should not be understood as that of
a discretionary authority, governance body, or punitive enforcer. Shiva denotes
the \emph{irreversible sink} required to couple continuation value with real
costs over time.

Shiva is external to the mechanism’s strategic domain but internal to the agents’
evaluative horizon. It is not a player, rule, layer, or governance body, and no
action within the mechanism can condition on it strategically. Yet agents
experience Shiva as time, irreversibility, and loss, and therefore incorporate
its effects directly into their private judgments of continuation value. In this
sense, Shiva supplies the functional role of authority without being an
institutional authority: it constrains behavior by rendering certain choices
irreversible, not by issuing commands, verifying actions, or enforcing rules.
Because Shiva cannot be manipulated, appealed to, or conditioned upon within
the mechanism, it remains authoritative only in the phenomenological sense that
its consequences are real to participants and unavoidable over time.

Hashing links two aspects of the same object:
\begin{enumerate}
\item \emph{Continuation value}, which agents trade and allocate through
strategic interaction.
\item \emph{Entropy expenditure}, which renders certain actions irreversible and
prevents the unwinding of exposure.
\end{enumerate}

These are not separable. Continuation value is only meaningful because it can be
irreversibly spent; entropy expenditure is only economically relevant because it
destroys continuation value. Framing Shiva as a ``high cost'' misunderstands its
function: it is not a deterrent priced in equilibrium, but the thermodynamic
constraint that makes time, commitment, and revision meaningful.

Without such an entropy sink, enforcement collapses into arbitrage, as exposure
can always be reversed or postponed. With it, costs are front-loaded under
uncertainty and cannot be neutralized by strategic play.

\subsection{Evaluation Criterion}

Incentive compatibility in these mechanisms must be evaluated over
\emph{histories and parallel games}, not reports. The relevant question is not
whether agents can profitably misreport preferences, but whether they can improve
expected welfare by manipulating the mechanism’s enforcement function without
bearing irreversible exposure.

The paper’s claims apply only under this evaluative frame. Within it,
impossibility results derived for direct mechanisms remain correct but
incomplete.
\clearpage

\end{document}